\documentclass[%
 aip,
 amsmath,amssymb,
 reprint,%
]{revtex4-2}

\usepackage{graphicx}
\usepackage{dcolumn}
\usepackage{bm}

\newcommand{\Rb}{$^{87}$Rb}
\newcommand{\ket}[1]{\left|#1\right\rangle}
\newcommand{\bra}[1]{\left\langle#1\right|}

\begin{document}


\title{Atomic microwave-to-optical signal transduction via magnetic-field coupling in a resonant microwave cavity}%

\author{A.\ Tretiakov}
\author{C.\ A.\ Potts}
\author{T.\ S. Lee}%
\author{M.\ J. Thiessen}%
\author{J.\ P.\ Davis}
\affiliation{%
Dept. of Physics, University of Alberta, Edmonton AB T6G 2E1
}%
\affiliation{%
Dept. of Physics, University of Alberta, Edmonton AB T6G 2E1
}%
\author{L.\ J.\ LeBlanc}
\email{lindsay.leblanc@ualberta.ca}
\affiliation{%
Dept. of Physics, University of Alberta, Edmonton AB T6G 2E1
}%

\date{\today}

\begin{abstract}

Atomic vapors offer many opportunities for manipulating electromagnetic signals across a broad range of the electromagnetic spectrum.  Here, a microwave signal with an audio-frequency modulation encodes information in an optical signal by exploiting an atomic microwave-to-optical double resonance, and  magnetic-field coupling that is amplified by a resonant high-Q microwave cavity. Using this approach, audio signals are encoded as amplitude or frequency modulations in a GHz carrier, transmitted through a cable or over free space, demodulated through cavity-enhanced atom-microwave interactions, and finally, optically detected to extract the original information.  This atom-cavity signal transduction technique provides a powerful means by which to transfer information between microwave and optical fields, all using a relatively simple experimental setup without active electronics. 

\end{abstract}


\maketitle


Manipulating and probing atoms with microwave-frequency radiation is a well-established technique. Despite the fact that the interactions between microwaves and atoms are weaker than optical interactions, microwaves play a leading role in applications ranging from atomic clock standards,\cite{Diddams2004} to cavity-QED-based quantum information studies,\cite{Haroche2013} to ac-magnetometry,\cite{Horsley2016} and to even trapping atoms and molecules.\cite{DeMille2004,Stammeier2018} As microwave-frequency photons become increasingly important across a variety of quantum information platforms, applications requiring enhanced coupling between microwaves and atoms have emerged, including quantum information transduction from microwave to optical frequencies,\cite{Hafezi2012,Reed2017,Han2018,Adwaith2018}
and quantum information storage in a microwave quantum memory.\cite{Kubo2012,Grezes2014,Probst2015,Wolfowicz2015}  

In the past decade, a new class of three-dimensional microwave cavities have emerged to enhance interactions with superconducting-circuit artificial atoms,\cite{Paik2011,Reagor2013,Kono2018,Lane2019} magnonic systems,\cite{Zhang2014d,Kostylev2016,Bai2017,Lachance-Quirion2017} and electro-mechanical resonators.\cite{DeLorenzo2014,Yuan2015c,Souris2017a,Menke2017}  Cavities with the same design principles can be used to enhance interactions between microwaves and real atoms.\cite{Sun2017,Adwaith2018}  In this work, we use a room-temperature tunable copper cavity resonant with the hyperfine resonance of the Rb atom, inside of which sits a glass vapor cell containing gas of Rb atoms. Microwave-field enhancement due to the cavity allows us to make use of the ground-state microwave coupling in an alkali-atom vapor, greatly simplifying our scheme as compared to recent ``radio-over-fiber'' (RoF) experiments with alkali vapors, where transmission of signals via amplitude (AM) and frequency (FM) modulation\cite{Deb2018,Anderson2018,Holloway2019e,Jiao2019} and digital communications\cite{Holloway2019d,Song2019,Meyer2018a} were successfully demonstrated. In those experiments,  information from the microwave (MW) field was transferred to a laser probe through electromagnetically-induced transparency in Rydberg states, modulated by the MW field via Autler-Townes splitting or by the ac Stark shift.

In this work, we apply a model that shows how, in a three-level double-resonance scheme similar to those used for power standards,\cite{Coffer2000}  ac~magnetometry,\cite{Sun2017,Sun2018} and atomic clocks,\cite{Bandi2012} any time-varying parameter in the coupling between atoms and one near-resonant field can be transduced to the second field nearly resonant to the other transition. Next, we demonstrate this principle in a simple experimental configuration using a cavity-enhanced microwave-atom coupling, and demonstrate phase-coherent microwave-to-optical signal transduction and information transmission.

\begin{figure}[tb!]  
\begin{center}
\includegraphics[width=85mm]{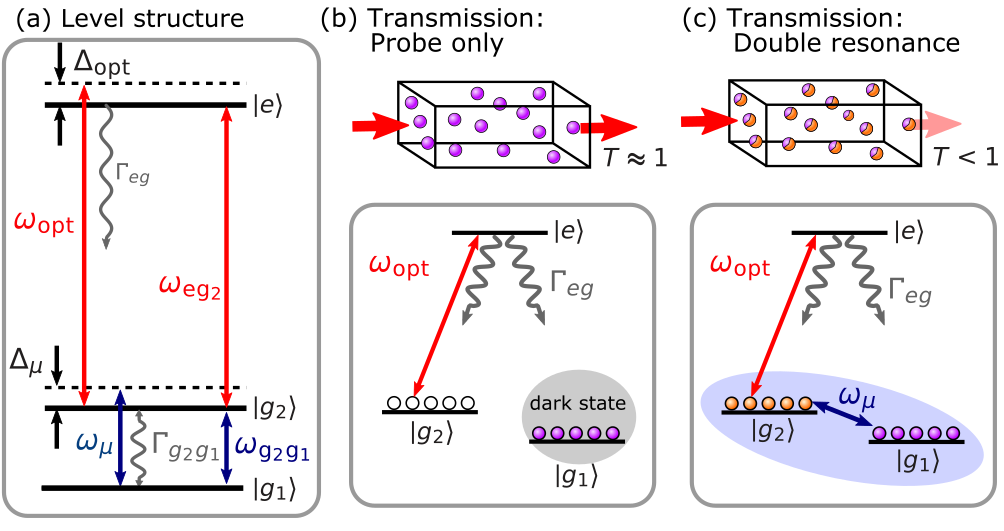}
\caption{Level structure and optical pumping. (a) Three-level structure, typical of an alkali atom-like \Rb, in which case $\ket{e} \rightarrow \ket{F^\prime =3}$ in the $^2P_{3/2}$ level and is connected optically to the two $^2S_{1/2}$ hyperfine ground states, $\ket{g_1} \rightarrow \ket{F =1}$ and  $\ket{g_2} \rightarrow \ket{F =2}$. Frequencies of the optical ($\omega_{\rm opt}$) and microwave ($\omega_{\mu}$) fields, as well as the relevant detunings, are indicated. (Energies not to scale.) (b) In the absence of microwave coupling, the atomic population is optically pumped to $\ket{g_1}$, a ``dark state.''  (c) With resonant microwave radiation, the system is in ``double resonance,'' and the transmission of the optical probe depends on the population in $\ket{g_2}$. }
\label{fig:pumping}
\end{center}
\end{figure}

The electronic structure of alkali atoms, like rubidium, includes readily accessible transitions in both the optical and microwave regime.  Here, we make use of a ``double-resonance'' scheme\cite{Demtroder1996} that involves an electric-dipole optical transition (at 384~THz, or 780~nm) and a magnetic-dipole transition in the ground-state hyperfine splitting of the \Rb~atom (6.8~GHz).
In our experiment, the microwave field couples the two hyperfine ground state levels $\ket{g_1}\equiv \ket{F =1}$ and $\ket{g_2} \equiv \ket{F = 2}$, and an optical probe connects  $\ket{g_2}$ to $\ket{e} \equiv \ket{F^\prime = 3}$  in the D$_2$-manifold of $^{87}$Rb  [Fig.~\ref{fig:pumping}(a)]. 

To understand the interplay between the optical and microwave signals, we first consider the case where the microwave field is absent. Here, the probe optically pumps the ground state atomic population to $\ket{g_1}$ (due to off-resonant excitation to the excited $F=2$ level), and the probe's absorption is minimized [Fig.~\ref{fig:pumping}(b)]. Upon applying a resonant or a near-resonant microwave field to the system, the ground state levels are mixed and some of the atomic amplitude resides in $\ket{g_2}$, which results in less transmission through the vapor [Fig.~\ref{fig:pumping}(c)]. The steady-state distribution of atomic populations between $\ket{g_1}$ and $\ket{g_2}$ depends on the values of the optical and microwave powers and detunings, and the spontaneous emission rates $\Gamma_{\rm eg}$, and thus, the proportion of probe transmission, $T$,  depends on all of these.  This spectroscopic technique is known as double-resonance imaging\cite{Bandi2012} and is the fundamental  mechanism exploited here.

\begin{figure*}
\includegraphics{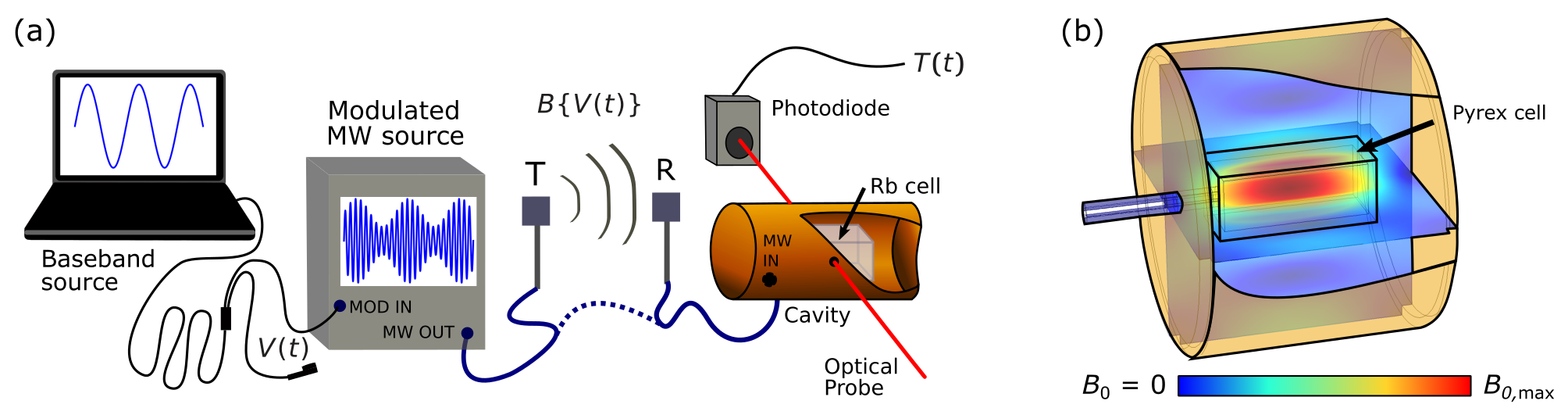}%
\caption{\label{fig:setup} (a) Experimental setup. An audio signal $V(t)$ is applied to the external modulation of a microwave (MW) frequency source. The modulated MW field transmitted via cable (solid) or antenna (dashed) to an MW cavity with a vapor cell inside. The atomic vapor transduces the modulation from the MW carrier to the optical probe intensity, which is converted to voltage via a photodiode. (b) Cavity field calculations, showing that the field of the TEM$_{011}$ mode is enhanced inside the Pyrex cell (outlined), due to its refractive index.}
\end{figure*}

More formally, these double-resonance dynamics can be described in terms of a three-level system [Fig.~\ref{fig:setup}(b)], plus external, time-dependent electric and magnetic fields $\mathbf{E}(t)$ and $\mathbf{B}(t)$ that are applied to the system. The coupled atom-fields Hamiltonian is 
\begin{align}
    \hat{H}=\hat{H_0}- \hat{\mathbf{d}} \cdot \mathbf{E}(t) -\hat{\boldsymbol{\mu}} \cdot \mathbf{B}(t),
\end{align}
where $\hat{H_0}$ is the atomic Hamiltonian, and $\hat{\mathbf{d}}$ and $\hat{\boldsymbol{\mu}}$ are the electric and magnetic dipole operators, respectively.  Due to the symmetries of the states involved, selection rules dictate that electric dipole transitions are allowed between states $\ket{g_{1,2}}$ and $\ket{e}$, but not between the ground states, and magnetic dipole transitions are allowed between $\ket{g_{1}}$ and $\ket{g_{2}}$, but not from these levels to $\ket{e}$. 

When dissipation is included,  the Lindblad equation
\begin{align}
    \dfrac{\partial\hat{\rho}}{\partial 
    t}=\dfrac{1}{i\hbar}[\hat{H},\hat{\rho}]+ L\{\hat{\rho}\}
\end{align}
more completely describes the system's dynamics, where $\hat{\rho}$ is the atomic density matrix, and $L\{\hat{\rho}\}$ is the Lindblad term,\cite{Lindblad} representing a phenomenological decoherence term that takes into account relaxation processes, such as spontaneous emission or collisional relaxation.

In the case of harmonic optical and microwave fields, where the optical electric field $\mathbf{E}(t) = \mathbf{E}_0\cos(\omega_{\rm opt} t +\phi_{\rm opt})$ is nearly resonant with the $\ket{g_{2}} \rightarrow \ket{e}$ transition and the microwave magnetic field $\mathbf{B}(t) = \mathbf{B}_0\cos(\omega_{\mu} t +\phi_{\mu})$ is nearly resonant with the $\ket{g_{1}} \rightarrow \ket{g_{2}}$ transition, the rotating-wave Hamiltonian (expressed in the basis $\{\ket{e}, \ket{g_2}, \ket{g_1}\}^{\rm T}$)  is
\begin{align}
    \hat{H}=\dfrac{\hbar}{2}\left( \begin{array}{ccc}{2\Delta_{\rm opt}} & {\Omega_{\rm opt}} & {0} \\ {\Omega_{\rm opt}} & {0} & {\Omega_{\mu}}  \\ {0} & {\Omega_{\mu} } & {2\Delta_{\mu}}\end{array}\right). 
\end{align}
Here,  $\Delta_{\rm opt} = \omega_{\rm opt} - \omega_{\rm eg_2}$ and $\Delta_{\mu} = \omega_{\mu} - \omega_{g_2g_1}$ are the optical and microwave detunings respectively, $\Omega_{\rm opt}=\bra{\rm e} \mathbf{d} \cdot \mathbf{E}_0 \ket{\rm g_2}/\hbar$ and $\Omega_{\mu}=\bra{\rm g_1} \hat{\boldsymbol{\mu}} \cdot \mathbf{B}_0 \ket{\rm g_2}/\hbar$ are the optical and microwave Rabi frequencies, and $\hbar$ is the reduced Planck's constant.  

After relaxation, the density matrix reaches a constant steady-state value that depends on the field parameters and decoherence rates. As described above, the transmission $T$ of the probe (which is determined by the population in level $\ket{g_2}$) will likewise reach its steady-state value. If any parameter (generically, ``$X$'') is subject to a time-dependent variation, such that $X\rightarrow X + \delta X(t)$,  the time-dependent transmission is
\begin{align}
T[X+\delta X(t)]\approx T(X)+\dfrac{\partial T}{\partial X}\delta X(t),  
\end{align}
where we assume that the variation is sufficiently small such that a first-order approximation is justified, and that any time-dependence is slower than a typical optical pumping time, to allow for quasi-steady-state behavior.

In the spirit of RoF applications, we consider how amplitude and frequency modulation of the microwave signal affects the optical transmission. The principal idea is similar to that of frequency-modulation spectroscopy\cite{Bjorklund1980}, except there, the change in the transmission is the result of the variation in the parameters of the probe itself.
For  amplitude modulation, the magnitude of the microwave field varies as ${B}_0 \rightarrow {B}_0[1+ m_{\rm AM} V(t)]$, where $V(t)$ is the voltage of the carrier signal, $m_{\rm AM}$ is the relative amplitude-modulation sensitivity. In the Hamiltonian, the affected term is $\Omega_{\mu}$, which is modulated in the same way: $\Omega_{\mu}(t)\rightarrow \Omega_{\mu,0}\left[1+m_{\rm AM}\cdot V(t)\right]$. In this case, the transmission varies in proportion to the  modulation $V(t)$,
\begin{align}
    T^{\rm AM}(t)\approx T_0+m_{\rm AM} V(t)\Omega_{\mu,0}\left(\dfrac{\partial T}{\partial \Omega_{\mu}}\right).
        \label{eq:TAM}
\end{align}

Similarly, for  frequency modulation, the frequency of the microwave field varies as $\omega_\mu \rightarrow \omega_{\mu,0} + m_{\rm FM} V(t)$, where $m_{\rm FM}$ is the frequency-modulation sensitivity. This time, the lowest-order affected term in the Hamiltonian is $\Delta_{\mu}$, which is modulated as: $\Delta_{\mu}(t)\rightarrow \Delta_{\mu,0}+m_{\rm FM} V(t)$. Here, the transmission varies as the  modulation $V(t)$:
\begin{align}
    T^{\rm FM}(t)\approx T_0+m_{\rm FM} V(t)\left(\dfrac{\partial T}{\partial \Delta_{\mu}}\right).
    \label{eq:TFM}
\end{align}
The dependence of ${\partial T}/{\partial \Omega_\mu}$ and ${\partial T}/{\partial \Delta_\mu}$ on the parameters of the MW field is disccussed below.



In our experiment, an audio source is connected to the modulation input of a 6.8~GHz frequency generator to effect amplitude or frequency modulation of a microwave carrier  [Fig.~\ref{fig:setup}(a)].  The MW signal is transmitted to a copper microwave cavity via SMA cable, or over free-space via antennas. The cavity surrounds a glass cell  connected to a vacuum system via a glass stem coming through the center of one the cavity's faces, and Rb vapor is supplied to the system using a commercial atomic source, which allows us to control the Rb vapor density. 

The microwave cavity used in these experiments localizes the microwave field in the central volume of the cavity, with polarization along the axis of cylindrical symmetry, in the $\rm{TE}_{011}$ mode.\cite{Reagor2013}  The high-quality copper and surface treatments result in a room-temperature quality factor $Q\approx 27,000$, even though holes were drilled in the cavity both across the body (as shown) to permit optical interrogation, and in one end-cap to allow the stem of the pyrex vapor cell to extend to an external rubidium source.  
This large value of Q, which leads to significant MW amplification, makes this setup plausible  for RoF applications, where MW amplitudes are typically low.

With the cross-body access holes, simultaneous optical- and microwave-field interactions are possible. Through double-resonance, the audio signal is mapped on to the transmitted optical probe power, which is then converted to a voltage by an amplified photodiode.

\begin{figure}[tb!]
\includegraphics{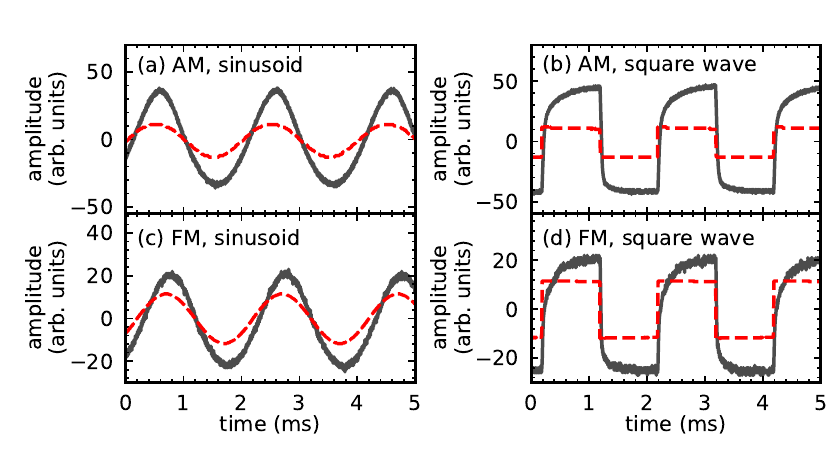}
\caption{\label{fig:characterization} Single-frequency  responses of optical transmission (grey) in the presence of microwave fields modulated at $\omega_{\rm m}/2\pi = 500$~Hz (red dashed) where input MW power corresponds to $\Omega_{\rm R}/2\pi = 74$~kHz.
(a) Sinusoidal and (b) square-wave  amplitude modulation with modulation sensitivity $m_{\rm AM} = 15\%/$V measured at $\Delta_\mu/2\pi = -5$~kHz. (c) Sinusoidal and (d) square-wave frequency modulation with modulation sensitivity of $m_{\rm FM} = 40$~kHz/V, measured at $\Delta_\mu/2\pi = 95$~kHz. AM and FM depictions are not to scale.}
\centering
\end{figure}

\begin{figure}[tb!]  
\begin{center}
\includegraphics[width=85mm]{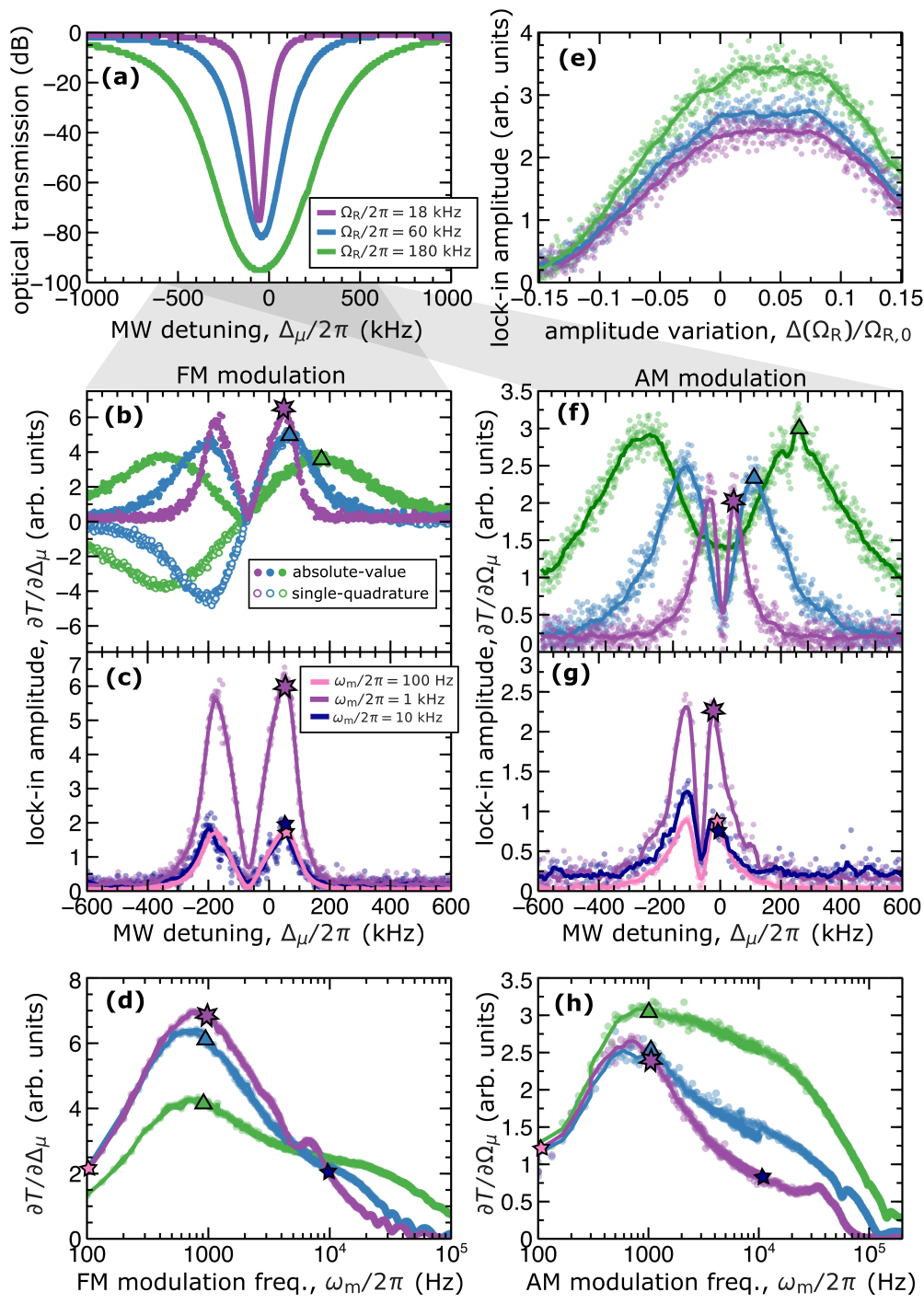}
\caption{Atomic radio calibrations, varying MW power [Rabi frequencies $\Omega_{\rm R}/2\pi = 18$~kHz (green), $60$~kHz (blue), and $180$~kHz (purple)] and modulation frequency for FM (b-d) and AM (e-h) signals. Throughout, the Rabi frequency relates to the input microwave power $P$  as $\Omega_{\rm R}/2\pi = 58.6~{\rm kHz} \times 10^{P{\rm[dBm]}/20}$. Symbols indicate identical conditions across subfigures within FM, AM categories. (a) Optical probe transmission vs.\ MW detuning; (b) lock-in detection signal amplitude for $1$-kHz-FM modulated signals, including absolute-value amplitude response (filled) and single-quadrature response (open);  (c) lock-in detection amplitude for $\Omega_{\rm R}/2\pi = 180$~kHz at FM modulation frequencies $\omega_{\rm m}/2\pi = 0.1$~kHz (pink), $1$~kHz (purple), and $10$~kHz (navy); (d) lock-in amplitude response vs FM modulation frequency, measured at the detuning with largest amplitude response. (e) Amplitude response vs change $\Delta(\Omega_{\rm R}$ from the nominal Rabi frequency, $\Omega_{\rm R,0}$ \; (f) absolute-value lock-in detection amplitude for $1$-kHz-AM modulated signals; (g) amplitude response, as in (c), for AM modulation when $\Omega_{\rm R}/2\pi = 180$~kHz; and (h) amplitude response, as in (d), for AM modulation.}
\label{fig:calibration}
\end{center}
\end{figure}

To demonstrate the initial proof-of-concept, single-frequency amplitude- and frequency-modulated microwave signals are applied, with modulation frequency $\omega_{\rm m}/2\pi = 500$~Hz.  Figures~\ref{fig:characterization}(a) and (b) show the probe transmission under AM for sinusoidal [$V(t) = V_0 \sin \omega_{\rm m}t$] and square-wave [$V(t) = V_0\rm{sgn}(\sin \omega_{\rm {m}}t)$] modulation signals, while Figs.~\ref{fig:characterization}(c) and (d) show the same for FM. As expected from Eqs.~(\ref{eq:TAM}) and (\ref{eq:TFM}), the transmission is, in all cases, proportional to the modulated signal $V(t)$, with a time-dependence evident in the ``roll-off'' of the square-wave signals that indicates the system's time response, as discussed below.

Next, we explore the system's dependence on various microwave parameters, including the detuning and power of the microwave carrier, and the frequency of harmonic modulation.  Along with a measure of the quasi-static (unmodulated) transmission of the optical probe in the presence of the unmodulated microwave carrier [Fig.~\ref{fig:calibration}(a)], Figs.~\ref{fig:calibration}(b,c) and (f,g) show ${\partial T}/{\partial \Delta_\mu}$ and ${\partial T}/{\partial \Omega_\mu}$
as a function of MW-carrier detuning $\Delta_{\mu}$ for different MW power and modulation frequencies measured via lock-in detection at $\omega_m$. 
Figures~\ref{fig:calibration}(d) and (h)  show the results of the same measurement, where the detuning is chosen to maximize the lock-in signal amplitudes, as a function of the modulation frequency.  

Square-wave modulation measurements, which show a typical response time of 1~ms [Fig.~\ref{fig:characterization}(b,d)], are consistent with the modulation frequency bandwidth seen in Figs.~\ref{fig:calibration}(d,h). In general, the response time is a manifestation of the finite-time for optical pumping that ``resets'' the quasi-steady state population distribution among the ground states, and this depends upon all parameters governing these dynamics. Additionally, like any driven oscillator, we expect a non-uniform response across frequencies, with a peak at some resonance, that  depends on the particular parameters of the system.~\cite{Sun2017,Tretiakov2019}

Even though the amplitude and bandwidth of the transduced signals have non-trivial dependence on several variables, the primary advantage of the current method  is that it is not necessary to know these parameters \emph{a priori}: so long as there is a dependence upon the relevant parameter (whether microwave amplitude or frequency, in this case) in the transmission, this method of signal transduction \emph{will} work.

\begin{figure}[tb!]
\includegraphics{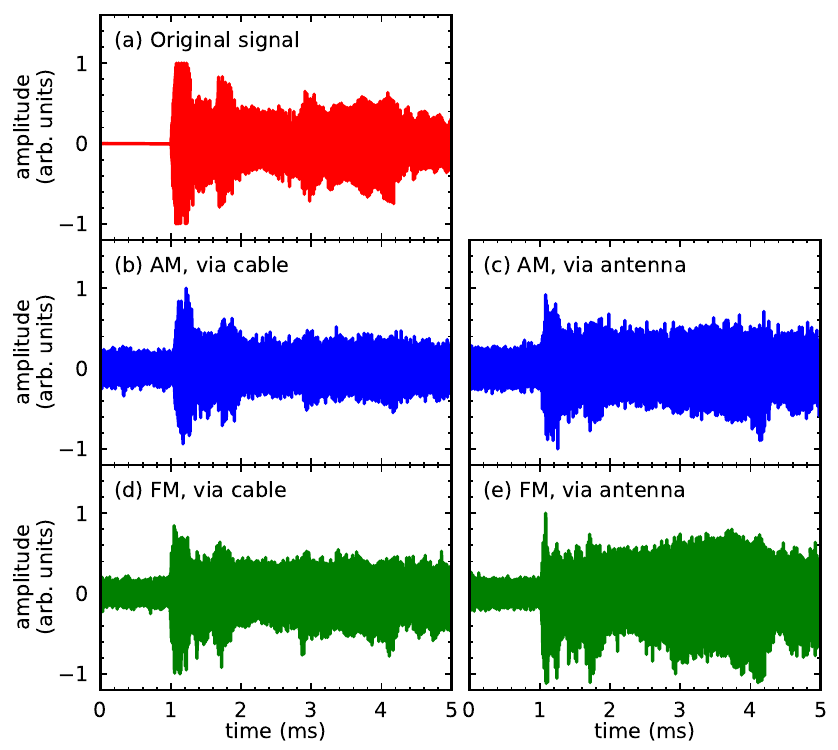}
\caption{\label{fig:results} Audio signal transduction. (a) Original audio signal $V(t)$ used to modulate the  microwave carrier $\omega_{\mu}$. (b, c) Optical transmission of amplitude modulated signal at $\omega_{\mu}/2\pi=6.834~682~610$-GHz [transmitted via cable (b) and 30-cm-separated antennae (c)] for MW input power $P=-5$~dBm and $m_{\rm AM}=15\%/V$. (d, e) Optical transmission of frequency modulated signal at $m_{\rm FM}=150$~kHz/V. (d) Signal transmitted over cable for MW input power  $P=-10$~dBm and $\omega_{\mu}/2\pi=6.834874610$-GHz. (e) Signal transmitted over 30-cm-separated antennae  for MW input power $P=+8 $~dBm and $\omega_{\mu}/2\pi=6.834877610$-GHz.  In all cases, a static bias magnetic field is applied to cancel stray magnetic fields.}
\end{figure}

Next, we demonstrate audio signal transduction using both AM and.  Using the principles discussed above, we encode an audio signal from a license-free song in the microwave field transmitted to the cavity input either directly (via cable) or over free-space between antennas.  The MW parameters were chosen to maximize the signal clarity and signal-to-noise ratio.
Figure~\ref{fig:results} and the supplementary audio files demonstrate AM and FM audio signal transmission.  We find that AM  more faithfully renders the signal, while the FM signal is more distorted, due to the asymmetric nature of the off-resonant sensitivity coefficient, as measured in Fig.~\ref{fig:calibration}(c).

In conclusion, we have demonstrated a powerful method for microwave-to-optical transduction of an analog audio signal based on cavity-enhanced magnetic-dipole coupling in a room-temperature alkali vapor. As a proof-of-concept, we have shown that this method is practical for AM and FM radio-over-fiber applications. An advantage of the setup as compared to other ``atomic radio'' systems is that cavity makes it possible to realize a configuration where the atoms do not need to ``receive'' the microwaves directly: the cavity-enhanced signal is strong enough to permit  indirect sensing via an antenna that is not rigidly attached to the cavity, which allows its arbitrary adjustment to optimize reception.

Though our demonstrations are bandwidth-limited by the system's finite response time in adjusting to  a new steady state, the agnosticism of this method with respect to the details of system parameters means that it will have broader applicability in other atom- and atom-like systems with faster responses.  For rubidium-atom applications, we foresee opportunities for digital signal encoding\cite{Song2019} and phase-shift keying,\cite{Rudd2019,Meyer2018a,Holloway2019d} both of which can be used to increase the information capacity.  Finally, we highlight that the high-Q room-temperature cavity made this demonstration possible, and look forward to future applications of room-temperature atomic systems with cavity-enhanced magnetic-dipole interactions, especially in the realm of quantum technologies.

\acknowledgements{This work was generously supported by the University of Alberta, Faculty of Science, the Natural Sciences and Engineering Research Council (NSERC), Canada (Grant Nos.\  RGPIN-06618-14, RGPIN-04523-16, DAS-492947-16, STPGP-494024-16, and CREATE-495446-17), Alberta Innovates, the Canada Foundation for Innovation, and the Canada Research Chairs (CRC) Program.}


\begin{thebibliography}{10}
\bibitem{Lindblad}
The Lindblad term is a general descriptor of relaxation processes,
  $L\{\hat{\rho}\}=\sum_{nm}\Gamma_{nm}\ket{m}\bra{n}\hat{\rho}\ket{n}\bra{m}
  -\tfrac{1}{2}\sum_{nm}\Gamma_{nm}\left(\hat{\rho}\ket{n}\bra{n}+\ket{n}\bra{n}\hat{\rho}
  \right)$.
\end{thebibliography}



\end{document}